\documentclass[aps,prd,twocolumn,preprintnumbers,nofootinbib,superscriptaddress,longbibliography]{revtex4-1}

\usepackage[utf8]{inputenc}
\usepackage{amsmath}
\usepackage{amssymb}
\usepackage{xspace}
\usepackage{xcolor}
\usepackage{color}
\usepackage{soul}
\usepackage{url}
\usepackage{appendix}
\usepackage{cancel}
\usepackage{graphicx}
\usepackage{subcaption}
\usepackage{fancyvrb}
\usepackage{setspace}
\usepackage{siunitx}
\usepackage{braket}
\usepackage{chemformula}
\usepackage[normalem]{ulem}
\usepackage[colorlinks=true,citecolor=blue,linkcolor=blue]{hyperref}

\usepackage[capitalise, english]{cleveref}
\crefformat{equation}{(#2#1#3)} 
\usepackage{bm}

\definecolor{red}{rgb}{0.9, 0,0}
\definecolor{cerulean}{rgb}{0., 0.62,0.9}
\definecolor{navy}{rgb}{0.05, 0.05,0.8}

\graphicspath{{./figures/}}

\newcommand{\bfq}{{\bf q}}

\begin{document}

\title{
Dark Matter Detection Using Phonon Sensing in Amorphous Materials
}

\affiliation{School of Physics, Peking University, Beijing 100871, China}

\author{Itay M. Bloch}
\affiliation{Physics Division, Lawrence Berkeley National Laboratory, Berkeley, CA 94720, USA}
\affiliation{Leinweber Center for Theoretical Physics, Department of Physics, University of California, Berkeley, CA 94720, USA}
\affiliation{Theoretical Physics Department, CERN, 1 Esplanade des Particules, CH-1211 Geneva 23, Switzerland}
\affiliation{Physics Department, Technion – Israel Institute of Technology, Haifa 3200003, Israel}

\author{Simon Knapen}
\affiliation{Physics Division, Lawrence Berkeley National Laboratory, Berkeley, CA 94720, USA}
\affiliation{Leinweber Center for Theoretical Physics, Department of Physics, University of California, Berkeley, CA 94720, USA}

\author{Xinran Li}
\email{xrl@pku.edu.cn}
\affiliation{School of Physics, Peking University, Beijing 100871, China}

\author{Amalia Madden}
\affiliation{Kavli Institute for Theoretical Physics, Santa Barbara, CA 93106, USA}

\author{Giacomo Marocco}
\affiliation{Physics Division, Lawrence Berkeley National Laboratory, Berkeley, CA 94720, USA}

\begin{abstract}

We present a concept for a tabletop-scale detector with an amorphous target designed to search for dark matter absorption into phonon excitations.
In crystalline materials, absorption occurs only at narrow resonances where the dark matter mass matches a zero momentum optical phonon mode, whereas amorphous targets provide a broadband response that can substantially enhance the absorption rate away from these resonances.
The predicted backgrounds arise from the relaxation of disorder-induced metastable defects in the amorphous target, as well as from low-energy noise intrinsic to superconducting phonon sensors. 
A prototype detector with a target mass of only a few  $\mu\mathrm{g}$ could provide broadband sensitivity to dark photon absorption across the 50 meV–200 meV mass range, probing 
up to two orders of magnitude beyond existing constraints. 

\end{abstract}

\maketitle

\section{Introduction}

Over the past decade, advances in low-threshold detector technologies have steadily pushed the reach of direct-detection dark matter experiments to smaller deposited energies. Dark matter absorption is a compelling target in this direction since the entire rest mass energy of the incident particle is deposited in the detector, producing a sharp monoenergetic signal. From a theoretical perspective, dark matter candidates with masses in the range $1\,\mathrm{meV} \lesssim m_\chi \lesssim 100\,\mathrm{keV}$ can be cosmologically long-lived while maintaining experimentally accessible absorption rates~\cite{Pospelov:2008jk}. Existing searches probe absorption at $m_\chi \gtrsim 1 \, \mathrm{eV}$ \cite{SENSEI:2023zdf,DAMIC:2019dcn}; reaching meV-scale sensitivity remains a key objective for next-generation direct-detection searches.

A simple model of dark matter with sub-eV mass is the dark photon, with a naturally light mass and small coupling to the Standard Model from kinetic mixing~\cite{Holdom:1985ag}. It may yield the observed dark matter relic density~\cite{Graham:2015rva, Agrawal:2018vin, Co:2018lka, Cyncynates:2023zwj, Cyncynates:2024yxm}, although the dark photon parameter space of minimal Abelian-Higgs models is constrained by the formation of vortices in the early universe~\cite{East:2022rsi}. While the proposed experiments BREAD \cite{BREAD:2021tpx} and DPHaSE \cite{Koppell:2025dmt} are projected to have broadband sensitivity in the meV-eV mass range, at present only a narrow band has been probed beyond solar constraints~\cite{An:2020bxd, XENON:2021qze} by the LAMPOST experiment~\cite{Baryakhtar:2018doz, Chiles:2021gxk}.

One approach to detecting dark photon dark matter is its absorption into phonon modes\footnote{High frequency vibrational modes in an amorphous material are highly localized and are therefore not phonon modes in the strict sense of the word. 
They however decay rapidly to modes that propagate, and in this paper we will therefore refer to both types of modes as ``phonons''. 
}, which has been studied extensively for crystalline targets \cite{Knapen:2017ekk,Griffin:2018bjn,Knapen:2021bwg,Mitridate:2023izi,Billard:2022cqd}. 
This process can be resonant when the dark matter mass matches the frequency of one of the optical phonons at the center of the Brillouin zone. 
If the dark matter mass does not match any of these narrow resonances, the absorption must instead occur through multiphonon processes. The rate for such excitations is suppressed by several orders of magnitude when compared with the single-phonon absorption~\cite{yu1996fundamentals}.
This means the dark matter absorption processes into phonons in crystals are effectively \emph{narrow-band}, with little opportunity to tune the resonance frequency, short of building many sensors with different materials or applying extreme external pressure to the target material~\cite{Ashour:2024xfp}. 

In this Letter, we point out that the lack of translation invariance in an amorphous solid implies a \emph{broadband} absorption response, which leverages the full spectrum of vibrational energy levels of the target. 
This implies that amorphous targets, such as glasses, can yield much larger dark matter absorption rates compared to (off-resonant) crystalline targets with similar exposures. 
The importance of disorder has long been appreciated in the condensed matter literature, for instance in the context of Raman spectroscopy and infrared absorption~\cite{shuker1970raman, smith1971raman, brodsky1974infrared, Tauc1974, adachi2012optical}. Its implications for dark matter detection, however, have yet to be fully explored.

Disorder has previously been exploited to relax the selection rules inherent to solid-state targets. For instance, it has been shown that for dark matter absorption on electrons, introducing dopants into semiconductor detectors can modify the detection threshold \cite{Du:2022dxf}. Additionally, randomly oriented nuclear spins in a crystalline target yield a broadband absorption for the axion–nucleus coupling \cite{Bloch:2024qqo}. More recently, it was shown that axion-to-photon and dark-photon-to-photon conversions in dielectric powders are also broadband processes \cite{Koppell:2025dmt}. A detailed experimental proposal was put forward for a dielectric powder haloscope equipped with a superconducting nanowire single-photon detector (SNSPD). While conceptually similar, in that case mesoscopic disorder arises by using powders made of varying sphere radii, while in our case amorphous materials are disordered at the microscopic site-level.

This Letter is organized as follows: In \cref{sec:absorption} we sketch the physical principles that explain the broadband absorption spectrum and present a potential detection scheme in \cref{sec:detector}. We conclude in \cref{sec:conclusions} and reserve further details on the absorption rate calculation, phonon propagation, detector design and background estimates for the appendices.

\section{Absorption in disordered materials\label{sec:absorption}}

The absorption rate of light dark matter in a condensed matter target depends sensitively on whether the target preserves discrete translational symmetry.
In crystals, this unbroken symmetry implies that the absorption must conserve momentum up to a reciprocal lattice vector.
To leading order in the small momentum transfer, the allowed transitions are therefore those for which the dark matter mass is resonant with a optical phonon at the origin of the Brillouin zone.
For other masses, absorption proceeds via multiphonon processes, whose rates are suppressed by several orders of magnitude relative to the resonant channel.

On the other hand, in a disordered material, the translation symmetry is fully broken. Therefore there does not need to be a mode in the material whose momentum exactly matches that of the dark matter.
This means that the dark matter can 
access transitions to all vibrational modes of the material, subject to energy conservation.
This effect also exists in crystals, if the dark matter interacts through spin-dependent couplings. In this case the disorder in the random orientations of the nuclear spin can break the translation symmetry and yield a  broadband absorption response~\cite{Bloch:2024qqo}.
In this Letter, we study the absorption of dark photon dark matter, which couples to the charge density in the material. To effectively break the translation symmetry in this case, we consider an amorphous material. 

\begin{figure}
    \centering
    \includegraphics[width=\linewidth]{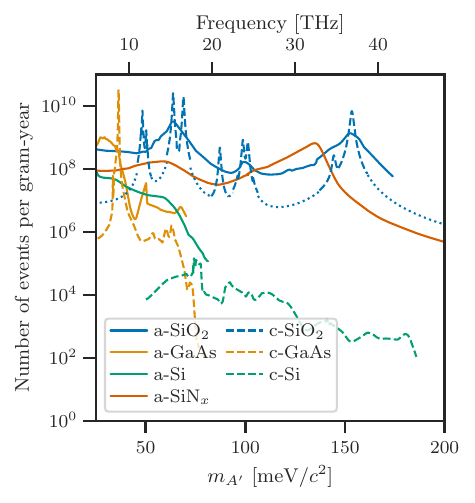}
    \caption{The maximal expected number of absorption events from dark photon dark matter for both amorphous (solid) and crystalline (dashed) targets, where at each $m_{A'}$ we take the dark photon mixing parameter $\kappa$ to saturate the upper limits derived from the XENON experiment~\cite{An:2020bxd, XENON:2021qze}. 
    For c-SiO$_2$, we indicate the regions more than 5 linewidths from the resonances by dotted lines, as the estimate of the excitation rate for these energies is merely an interpolation. See \cref{app:ELFs} for more details and references.
    }
    \label{fig:fiducial_n_events}
\end{figure}

We consider a massive dark photon $A'$ that couples to the electromagnetic current $J_\mathrm{EM}$ via the interaction
\begin{align}
    \mathcal{L}_\mathrm{int} = e \kappa  J_\mathrm{EM}^{\mu} A'_{\mu},
\end{align}
in the basis in which the photon and dark photon mass and kinetic terms are diagonal. The absorption rate per unit mass for dark photon dark matter is given by~\cite{Knapen:2021bwg,Knapen:2017ekk,Hochberg:2016ajh,An:2013yua,An:2014twa,Hochberg:2016sqx} 
\begin{equation}
R = \frac{\rho_{A'}}{\rho_\mathrm{T}}\kappa^2 \text{Im}\left[\frac{-1}{\epsilon(\omega)}\right]c^2/\hbar,
\end{equation}
with $\kappa$ the kinetic mixing parameter for a dark photon with mass $m_{A'}$. The local dark matter mass density is denoted by $\rho_{A'}\approx0.4\, \text{GeV}/\text{cm}^3$, while
$\rho_\mathrm{T}$ is the mass density of the target and $\epsilon(\omega)$ its complex dielectric function, evaluated at the frequency corresponding to the dark photon mass $ \omega=m_{A'} c^2$. The dark photon absorption spectrum is proportional to that of infrared photons, which enables us to extract the loss function $\text{Im}\left[\frac{-1}{\epsilon(\omega)}\right]$ from measurements
or from density functional theory computations. 
At the energy range of interest, 
all absorbed energy is expected to be dissipated into vibrational modes.

\cref{fig:fiducial_n_events} shows the dark matter absorption for several crystalline and amorphous targets, where for each $m_\chi$ we have chosen the largest value of $\kappa$ that is compatible with current experimental bounds~\cite{An:2020bxd, XENON:2021qze}. 
As can be seen, amorphous materials have a far broader response than their crystalline analogues, due to the disorder-induced weakening of the crystal selection rules.
In crystalline materials, these selection rules imply that off-resonant absorption must occur through multiphonon processes, which are heavily suppressed.
The absorption rate in apolar crystals such as Si is suppressed even further~\cite{brodsky1974infrared}.

\section{Detection concept\label{sec:detector}}

From an experimental point of view, there are two major differences between amorphous and crystalline targets: Firstly, in a very cold amorphous target, the phonons do not propagate ballistically as they do in a crystal, except for extremely low-energy phonons. This complicates the readout and restricts the size of the target. Secondly, the high degree of disorder implies that there is a large density of metastable states. These states, so-called ``two-level systems'' (TLS), can remain populated at low temperatures, and their relaxation during an experiment is an important background.
In this section, we provide an example detector design and discuss the expected background rates. 

\subsection{Example detector design}

\begin{figure}
    \centering
    \includegraphics[width=\linewidth]{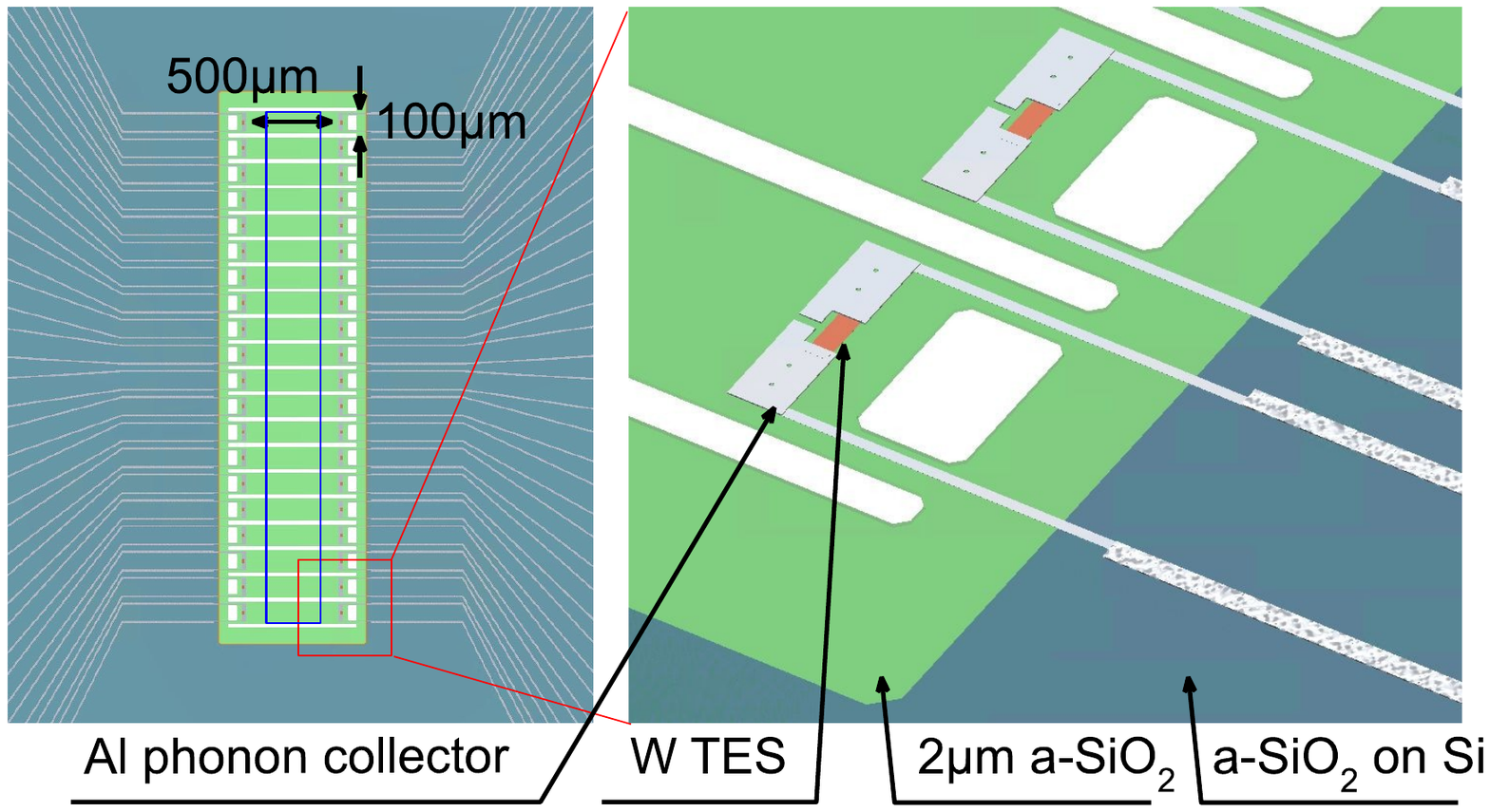}
    \caption{Conceptual design of the amorphous detector. The detector consists of thin amorphous dielectric membranes (green) suspended on the top of a silicon frame (blue). The membrane is etched into strips and the signals are read out by the two superconducting sensors on the two ends of the strips, TESs are drawn as an example. 
    The white areas are openings in the membrane, separating the strips from one another, to improve phonon collection efficiency. 
    The fiducial target volume consists of the middle region of the strips (highlighted by the blue box), as detailed in \cref{app:detector}\label{fig:conceptualDesign}.}
    
\end{figure}

 A possible design is to fabricate sensitive superconducting sensors on amorphous dielectric membranes, such as $\text{SiN}_x$ or $\text{SiO}_2$, with subscript $x$ indicating the non-stoichiometric form typically realized in thin-film fabrication. The membranes are suspended on a silicon frame, as shown in 
 \cref{fig:conceptualDesign}, and the membrane is further etched into strips which are individual detector channels. Potential DM signals are read out from the superconducting sensors fabricated on the two ends of the strips. 

The superconducting phonon sensors are made out of a Transition Edge Sensor (TES)~\cite{irwin2005transition} or kinetic inductance detector (KID)~\cite{zmuidzinas2012superconducting}, coupled to aluminum phonon collectors. The aluminium phonon collectors have an inherent energy threshold of 0.36 meV, which is the energy needed to break a Cooper pair in the Al. At this energy, the phonon mean free path in amorphous $\text{SiO}_2$ is only $\lambda \approx 10\, \mu\text{m}$. For any region of the target that is further away than $\sim \lambda$ from the sensor, the propagation is therefore diffusive rather than ballistic. (See \cref{app:diffusion}.) 

We further estimate that at these energies, the initial phonons decay rapidly, producing secondary phonons that can propagate up to $\sim 1\,\mathrm{mm}$ before reaching energies below the $0.36\,\mathrm{meV}$ threshold.
Consequently, detecting athermal phonons in an amorphous solid requires the sample to be considerably smaller than would be necessary for crystalline targets. 
This is also why we consider a target in the form of a thin membrane, as this effectively means that the phonons diffuse only in two dimensions, thus increasing the collection efficiency of the sensors on the surface.
In the design shown in \cref{fig:conceptualDesign}, the lengths of the strips are chosen to be the maximum that allows more than $90\%$ of the targeted energies to be collected by the sensors based on a simplified simulation of phonon diffusion (see \cref{app:detector}). 
A smaller target also allows for better energy resolution, as discussed below.

The sensitivity of the superconducting sensors scales inversely with square root of their volume \cite{irwin2005transition, zmuidzinas2012superconducting}; consequently, achieving a sub-\SI{100}{\milli\electronvolt} energy threshold requires extremely small volumes. In TES-based phonon sensors, high phonon collection efficiency can nevertheless be achieved by exploiting the quasiparticle-trapping technique~\cite{irwin1995quasiparticle, saab2000design, chang2025spontaneous}. This is accomplished with the Al phonon collectors, as shown in \cref{fig:conceptualDesign}.

While this design enables excellent single-channel performance, the maximum size of an individual target strip remains highly constrained, motivating the use of multi-channel arrays to achieve a larger total target mass. In a future generation of such an experiment where large-scale multiplexing is required or advantageous, KIDs could provide an attractive alternative to a TES readout due to their intrinsic frequency-domain multiplexability~\cite{zmuidzinas2012superconducting}. In such an implementation, the KID capacitor can be fabricated directly on the silicon frame, where the amorphous dielectric layer is etched away. This placement reduces TLS-induced phase noise, while routing the inductor onto the suspended membrane allows it to efficiently collect athermal phonons. 

Both TESs and KIDs have been fabricated on suspended SiN membranes~\cite{irwin2005transition, agrawal2021strong} using techniques such as deep reactive ion etching and $\text{XeF}_2$ etching. We therefore expect that the microfabrication of the conceptual design described above is feasible.
Single-photon counting at photon energies as low as $50$ meV has already been demonstrated in KIDs~\cite{day202425}, showing that extremely low energy thresholds are achievable. In addition, novel superconducting qubit-based sensors with meV sensitivity are being actively developed~\cite{Magoon:2026jqt, ramanathan2024quantum}. 

\subsection{Detector resolution and background} 
The sensor resolution
will be limited by the shot noise from high-rate
phonon bursts below the threshold. 
This includes TLS relaxation in the amorphous target and the low energy excess (LEE) observed in current cryogenic phonon detectors.
Thermal fluctuation noise in TESs or generation-recombination noise in KIDs will be subdominant in comparison because of the small volume of the sensor. 

So far, two components of the LEE that contribute to sensor resolution have been identified~\cite{chang2025spontaneous}. The first is a component localized to each TES channel, which is likely due to stress in the superconducting films. This component is therefore expected to be present in the proposed detector, as it employs similar materials. The sensor area in each channel is reduced such that the contribution to the resolution resulting from shot noise remains below 10 meV.

The second LEE component originates from the bulk of crystalline targets and its origin is unknown. Because the proposed detector has a much smaller volume, the bulk LEE contribution becomes negligible, even if the unknown source is also present in amorphous materials.\footnote{There is some evidence that the LEE is at least in part due to relaxation of microscopic, internal stress in the sample \cite{Anthony-Petersen:2022ujw}. 
In amorphous materials such relaxation processes involving small, localized of number of atoms are modeled by the relaxation of TLS.
To estimate the expected LEE in the amorphous sensor, we rescale the LEE rate as measured in crystals and add the TLS relaxation rate. 
It is therefore possible that part of the TLS relaxation rate is already accounted for in the rescaled LEE rate as measured in crystals.
} 
Similarly, due to the small detector volume, the TLS-induced shot noise is expected to be negligible. 

In practice, the resolution is fundamentally determined by the Fano noise of quasiparticle creation in Al phonon collectors, which scales as $\sigma_\mathrm{Fano} = \sqrt{F2\Delta_\mathrm{Al}E/\eta}$, where $F$ is the Fano factor and $\eta\approx0.5$ is the pair-breaking energy efficiency. 
The Fano factor quantifies the degree of correlation in the pair-breaking process during the signal energy deposition;
we assume $F=1$, corresponding to Poissonian noise. These choices imply a $10$ meV energy for a 100 meV energy injection.
Based on this, we assume that the trigger threshold can be set to $50$ meV, 5 times the resolution, and refer to \cref{app:detector} for further details and figures.

\begin{figure}
    \centering
    \includegraphics[width=0.95\linewidth]{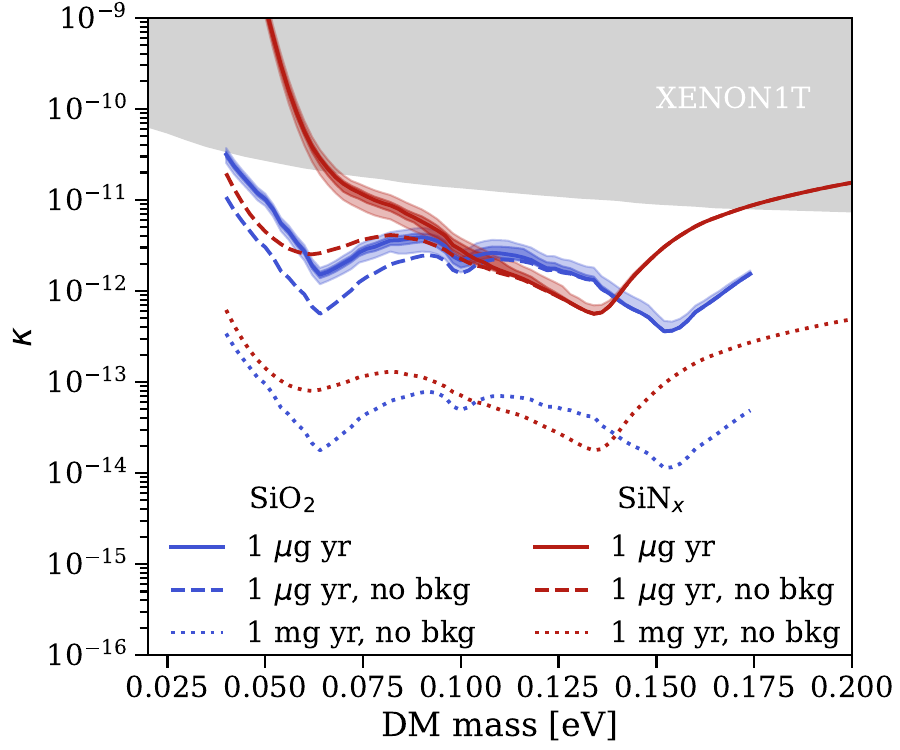}
    \caption{Exclusion sensitivity of the conceptual detector, along with existing bounds (gray shading) \cite{An:2020bxd, XENON:2021qze}. The solid lines are median sensitivity assuming the TLS background rate as calculated in \cref{app:background}. The dark and light shadow bands indicate $10\%,~30\%,~70\%$, and $90\%$ percentiles of the sensitivity projection. 
    The dashed (dotted) lines are ideal sensitivity assuming no backgrounds with \SI{1}{\micro\gram\,yr} (\SI{1}{\milli\gram\, yr}) exposure; they are not expected to be attainable and included only to illustrate the effect of the background on the expected sensitivity.
    The detector resolution is assumed to be dominated by the quasiparticle Fano noise, as quantified in \cref{app:detector}.
   }
    \label{fig:projected_sensitivity}
\end{figure}

Above the threshold, the background from TLS relaxation can be important: As the experiment is cooled down to cryogenic temperatures, a population of the TLS is initially in their excited state. 
They subsequently relax through quantum tunneling during the data taking period, injecting unwanted phonons into the detector.
Most, but not all, TLS decay quickly enough that we can simply wait for them to disappear before the data taking.
The remaining TLS relaxation rate depends strongly on their characteristic frequency, which is material dependent (see \cref{app:background}). 
We expect the TLS relaxation rate to produce a significant background rate in some materials (e.g. in a-SiN$_x$), while remaining negligible above the energy-threshold in others (e.g. in a-SiO$_2$).

The LEE component that is not due to TLS relaxation in the target is highly suppressed due to the small volume of the sensor, but the LEE component from the superconducting film remains significant. However, the two-channel readout design in \cref{fig:conceptualDesign} can reject this background by reconstructing the event position along the strip with the energy fraction in the two channels (see \cref{app:detector}).

The expected exclusion sensitivity is shown in \cref{fig:projected_sensitivity} and includes our estimates for the backgrounds as described above.
We also included a ``background-free'' projection (dashed lines), to illustrate how the background impacts the sensitivity.
In the case of a SiO$_2$ detector, the background rate in most of the signal region is expected to be low, and therefore it is worthwhile running a one-year exposure experiment to search for $m_{A'}\gtrsim$ 40 meV.
The sensitivity on $\kappa$ scales with the square root of the exposure as long as the expected number of background events is $\lesssim 1$ for a given mass hypothesis.
A SiN$_x$ detector is expected to be free of backgrounds for $m_{A'}\gtrsim 100$ meV; for lower $m_{A'}$ the TLS background is large and the projected sensitivity in this region is therefore weaker than for SiO$_2$. 
Here the sensitivity on $\kappa$ scales with the fourth root of the exposure.

\section{Discussion\label{sec:conclusions}}We have presented a concept for a phonon sensor that searches for dark photon dark matter using an amorphous target. Relative to crystalline materials, amorphous targets offer an inherently broadband absorption rate, allowing them to exceed the absorption of comparable crystalline detectors by one to two orders of magnitude over much of the 50 meV to 200 meV mass range.

A key practical distinction relative to crystalline targets is that athermal phonons in amorphous materials propagate diffusively rather than ballistically, favoring detector geometries based on thin membranes. Although this limits the size of individual targets, the reduced volume may be advantageous for background mitigation: shot noise from sub-threshold bulk processes scales with volume, which may lead to improved energy resolution in smaller sensors. This could both lower the detection threshold and narrow the energy-bin width, which is particularly valuable for signals that are mono-energetic, such as dark matter absorption.
This further enables a robust discovery strategy: with sufficient exposure, a genuine signal would emerge above background fluctuations, and observing a resonance at the same energy in multiple distinct materials would provide compelling evidence for a source external to the detector itself. 

Athermal phonon sensors are known to exhibit a low-energy excess (LEE) whose origin is not fully understood~\cite{adari2022excess, Baxter:2025odk}. In crystalline detectors, this excess has been attributed in part to stress relaxation through microfractures and to the diffusion of defects and impurities~\cite{Astrom:2005zk, Anthony-Petersen:2022ujw, Anthony-Petersen:2024vdh}. In projecting backgrounds for amorphous targets, we rescaled these empirically observed crystalline LEE contributions and additionally included a contribution from the relaxation of disorder-induced TLSs, which is a close amorphous analogue of stress- and defect-driven relaxation processes in crystals. 
It is possible that an amorphous detector may provide further insight into the physics behind the LEE itself, by comparing the background rates with those in analogous crystal-based sensors.

In addition, TLSs in amorphous oxide materials are a well-known source of decoherence in superconducting qubit devices~\cite{muller2019towards}. Therefore, we expect the data collected by this type of detector to have an impact beyond particle physics fundamental research: by enabling spectroscopy of TLS relaxation over a broad energy range, the detector concept presented here could inform and constrain microscopic models of TLS relevant for qubit-based quantum technologies.

Given our estimated background rate, we project that a single microgram-scale detector could already access new regions of dark photon parameter space. The background is dominated by the relaxation of TLSs. We estimated this rate from the phenomenological model of ~\cite{PhysRevB.84.174109}, the parameters of which were fit to data on ultrasonic dissipation (see \cref{app:background}).  Looking ahead, if our background estimations are borne out in a pilot detector, multiple detectors could be multiplexed to enable a scalable increase in effective target mass. 
For materials such as amorphous SiO$_2$ and SiN$_x$ this may allow for an experiment with higher exposure which could gain sensitivity to smaller couplings.
Should an excess be observed, one could attempt to further improve the sensitivity by building a crystalline detector with a resonance frequency at the observed excess.
More generally, the intrinsically broadband response of amorphous targets may be of benefit in searches for the absorption of any ambient field that couples to nuclei, including high-frequency gravitational waves~\cite{Kahn:2023mrj} and light scalars with dilaton-like couplings~\cite{Mitridate:2023izi}.

\section*{Acknowledgements}
We thank Daniel Carney, Maurice Garcia-Sciveres, Junwu Huang, Hugh Lippincott, Tongyan Lin, Ben Mazin, Nick Rodd, Brooke Russell, Matt Pyle,  Chiara Salemi,  Aritoki Suzuki and Alexandre Homrich for useful discussions. 
We particularly thank Maurice Garcia-Sciveres, Tongyan Lin and Aritoki Suzuki for comments on the manuscript.
AM was supported by grant GBMF7392 from the Gordon and Betty Moore Foundation and NSF PHY-2309135 to the Kavli Institute for Theoretical Physics (KITP). AM gratefully acknowledges the Pacific Postdoctoral Program at the Dark Universe Science Center, University of Washington, where part of this work was carried out. 
The Pacific Postdoctoral Program is supported by a grant from the Simons Foundation (SFI-MPS-T-Institutes-00012000, ML).
This work is supported in part by the Office of High Energy Physics of the U.S.\ Department of Energy under contract DE-AC02-05CH11231. 

\bibliography{refs.bib}

\appendix 

\section{Target materials}
\label{app:ELFs}

In this section, we provide the source of the data used to construct the energy loss function $\text{Im}\left[\frac{-1}{\epsilon}\right]$, which are listed in table~\ref{tab:materials}. We further provide details of any analytic approximations we use in calculating $\epsilon$.

Due to its resonant behavior, the crystalline response depends sensitively on the linewidth of the phonons. However, the linewidth broadens in the presence of phonons in the crystal, and is therefore temperature-dependent.
Since a realistic experiment would be run at cryogenic temperatures, we therefore require a determination of $\epsilon(\omega)$ either from low-temperature measurements or from a dedicated calculation of the multiphonon response using density functional theory methods. 
For GaAs, such a low temperature calculation was carried out in \cite{PhysRevB.70.245209}. The TESSERACT collaboration has also extensively studied $\text{SiO}_2$ and $\text{SiN}$ but we are not aware of low temperature calculations or measurements for these materials. 
For $\text{SiO}_2$ we therefore resort to the analytic approximation in \cite{Griffin:2018bjn}. 

For crystalline silica in particular, we use the analytic formula 
\begin{align}
\epsilon(\omega)=\epsilon_{\infty} \prod_\nu \frac{\omega_{\mathrm{LO}, \nu}^2-\omega^2-i \omega \gamma_{\mathrm{LO}, \nu}}{\omega_{\mathrm{TO}, \nu}^2-\omega^2-i \omega \gamma_{\mathrm{TO}, \nu}},
\label{eqn:silicaApproximation}
\end{align}
with the phonon frequencies taken from~\cite{gervais1975temperature}. Given the lack of data at the relevant cryogenic temperatures, we assume all the modes have a Q-value of 100, which is of the order-of-magnitude of those found in~\cite{gervais1975temperature}. The sensitivity curves involving crystalline silica should therefore only be interpreted as illustrative in the absence of precise knowledge of the phonon width. Additionally, far off-resonance (indicated by the dotted blue lines on \cref{fig:fiducial_n_events}), it is unclear how well the interpolation provided by equation~\ref{eqn:silicaApproximation} approximates the true dielectric response, as multi-phonon contributions are dominant here. 

\begin{table}[h!]
\centering
\begin{tabular}{l|l} 

  &  Reference \\
 \hline
 a-Si & \cite{brodsky1974infrared, palik1998handbook}  \\
 c-Si & \cite{ikezawa1981far, Knapen:2021bwg} \\
 a-SiO$_2$ & \cite{koike1989optical} \\
 c-SiO$_2$ & \cite{gervais1975temperature}* \\
 c-GaAs & \cite{PhysRevB.70.245209, palik1997gallium}  \\
 a-GaAs & \cite{stimets1973far, prettl1973far} \\
  a-SiN$_x$ & \cite{ cataldo2012infrared} 

 \end{tabular}
 \caption{A list of the materials considered, including both amorphous (a-) and crystalline (c-) targets. *: For crystalline silica, we use measurement of the phonon mode frequencies with the analytic equation 
 \eqref{eqn:silicaApproximation}.}
 \label{tab:materials}
 \end{table}

\section{Dynamics of vibrational modes in cold glasses \label{app:diffusion}}
The vibrational modes in an amorphous material differ fundamentally from those in a crystal, due to the lack of long-range order. Their properties can be studied experimentally and through many-body simulations; see \cite{Allen01011999} for a comprehensive review.
In this appendix, we review the features relevant to the direct detection of dark matter.

In bulk crystals, low frequency phonons can travel quasi-ballistically and their mean free path is set by the defect concentration.
In materials with a lot of disorder, the vibrational eigenmodes are in general not well approximated by plane waves, and it is much more difficult for those modes to propagate.
At high frequencies ($\omega \gtrsim 80$ meV), this effect is strong enough for the vibrational modes to be exponentially localized.
Modes with intermediate frequencies ($10\,\mathrm{meV} \lesssim\omega \lesssim 80$ meV) have a mean free path that is shorter than their wavelength and therefore cannot be thought of as a ballistically propagating wave packet, even on microscopic scales. In other words, they do not have a well-defined momentum vector and instead their wave function spreads out through quantum diffusion.
For low-frequency modes ($\omega \lesssim 10$ meV), the mean free path does exceed the wavelength of the perturbation, which means that a meaningful wave vector can be defined. 
One can think of these modes as propagating ballistically for at least a few times the typical inter-atomic distance. The corresponding mean free path is however orders of magnitude shorter than that of an acoustic phonon with the same frequency in a cold crystal.
The long-distance propagation of low frequency phonons in amorphous materials is therefore \emph{diffusive}\footnote{A priori, the evolution of the wavefunction of these modes is governed by many-body quantum mechanics. However, at low $\omega$ and in the limit of many scatterings, the equations simplify and the spatial probability density of these low-frequency modes satisfies the classical diffusion equation \cite{Akkermans_Montambaux_2007}. As we always work in that regime, we may treat the propagation of these phonons as diffusive in the classical sense. }, in contrast to the ballistic propagation of acoustic phonons in crystals. 

The lifetime of the vibrational modes can be estimated via numerical simulations \cite{PhysRevB.67.224302,PhysRevLett.77.3839}. For $\omega \gtrsim 10$ meV, it is found to be $\mathcal{O}(\mathrm{ps})$ or shorter. This lifetime is comparable to that of optical phonons in a crystal. 
The simulations moreover show that the decay rate is proportional to the two-mode density of states, consistent with the decay of a vibrational mode into two vibrational modes with lower frequencies.
The lifetime has not yet been simulated for $\omega \lesssim 10$ meV due to computational limitations.
In that regime the modes should already have a well-defined momentum vector and dispersion relation, though still with a mean-free-path that is much shorter than the detector's size.
We therefore extrapolate the lifetime to lower frequencies with the $\tau \sim \omega^{-5}$ scaling that is inherent to the Debye density of states.

\begin{figure}
    \centering
    \includegraphics[width=\linewidth]{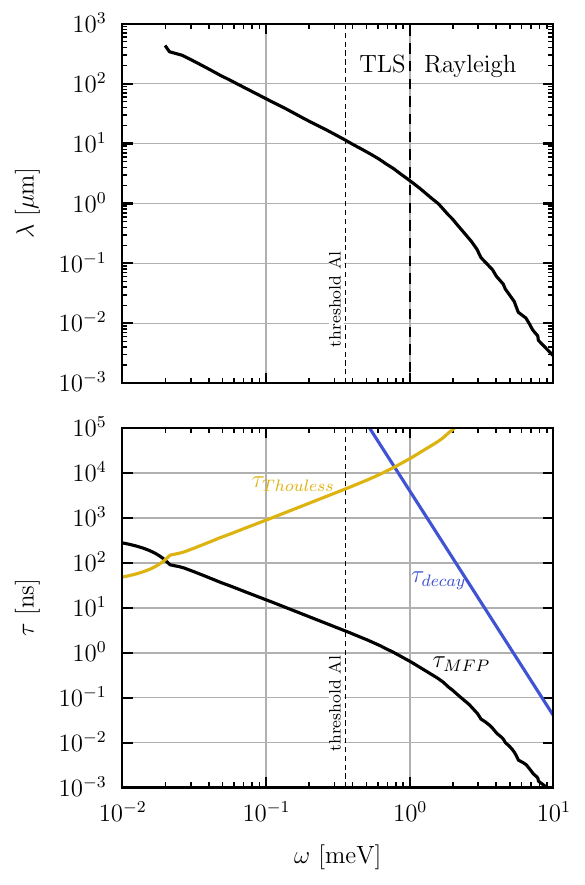}
    \caption{\textbf{Top:} Mean free path as a function of vibration energy for amorphous $\text{SiO}_2$ \cite{PhysRevB.4.2029}. The vertical dashed line indicates the energy threshold for a phonon to break a Cooper pair in the Al phonon collectors. The thicker vertical dashed line corresponds to transition between Rayleigh- and TLS-dominated scattering. \textbf{Bottom:} Characteristic timescales in a thin film of amorphous $\text{SiO}_2$. The Thouless time is computed for the distance from the center of the strip to the phonon collectors ($L\approx 250\,\mu$m).}
\label{fig:meanfreepath}
\end{figure}

As a practical matter, this means that all high-energy modes decay rapidly to lower-energy modes that are able to propagate, similar to the decay of high-frequency athermal phonons in a crystal. The key difference is that the propagation is diffusive rather than ballistic, which limits the volume of the target with a realistic read-out scheme.

Zeller and Pohl used measurements of the thermal conductivity as a function of temperature to extract the frequency dependence of the mean free path in the dominant phonon approximation \cite{PhysRevB.4.2029}. 
In this approximation, one assumes that the heat is conducted exclusively by the phonon frequency that provides the largest contribution to the specific heat, as a function of temperature. 
Further assuming a Debye model, this results in the conversion factor 
\begin{equation}
\omega \approx 0.37\,\mathrm{meV}\times \frac{T}{1\,\text{K}}
\end{equation}
Zeller and Pohl estimate the possible offset associated with this approximation to be less than a factor of 2 \cite{PhysRevB.4.2029}.

The resulting mean free path ($\lambda$) for amorphous $\text{SiO}_2$ is shown in the top panel of \cref{fig:meanfreepath}.
In the range $1\, \text{meV}\lesssim \omega \lesssim 10\,\text{meV}$, they observe an approximate $\omega^{-4}$ powerlaw, which Zeller and Pohl explain by Rayleigh scattering. 
For $\omega \lesssim 1\,\text{meV}$, the mean free path scales approximately as $\sim \omega^{-1}$, which is due to interactions with a class of defects called ``two-level systems'' (TLS). We will return to the TLS shortly, as they contribute to the expected background rate in the detector.
For $\omega \gtrsim 10\,\text{meV}$, not shown here, the attenuation of the modes is dominated by rapid anharmonic decays to lower frequency modes. 
 
In the bottom panel of \cref{fig:meanfreepath}, we compare the characteristic time scales relevant for the propagation of the vibrational mode through an amorphous material.
$\tau_{\mathrm{MFP}}$ is the time scale between interactions, which is the mean free path divided by the sound velocity, which is \mbox{$v_s\approx0.37 \mu$m/ns} for amorphous $\text{SiO}_2$.
$\tau_{\mathrm{decay}}$ is the time scale at which the excitations decay to two lower frequency excitations through anharmonic interactions, as explained above. 
Finally, the $\tau_{\mathrm{Thouless}}$ is the Thouless time, which is the expected timescale for the mode to diffuse to the surface of the sample. 
It is given by
\begin{equation}
\tau_{\mathrm{Thouless}} = \frac{L^2}{\frac{1}{3} v_s \lambda}
\end{equation}
with $L$ the most relevant dimension of the sample. In our case this is typical distance from a dark matter event in the center of the detector to the phonon collectors at the edge of the strips ($L\approx250\,\mu$m).
We see that high frequency modes continue to decay as they diffuse throughout the sample, until the $\tau_{\mathrm{Thouless}} < \tau_{\mathrm{decay}}$. 
For these frequencies, the mode is likely to reach the surface before it has a chance to decay. The $\tau_{\mathrm{Thouless}} \approx \tau_{\mathrm{decay}}$ crossover point therefore set the characteristic energy of the modes that reach the instrumented area of the sensor. 
For very low energies, one enters the regime where $\tau_{\mathrm{Thouless}} < \tau_{\mathrm{MFP}}$. 
Those modes effectively travel ballistically on the scale of the size of the sample. 

\section{Additional details on detector design\label{app:detector}}

\subsection{Benchmark detector configuration}

The conceptual design of the detector reflects the diffusive, rather than ballistic, propagation of phonons, which limits the size of the target.
In addition, the design should mitigate the following three background sources:
\begin{enumerate}
\item TLS decays:  This background source is specific to amorphous materials and originates from the bulk of the detector. We estimate this rate in \cref{app:background}.

\item Low energy excess (LEE) events in bulk of the detector \cite{chang2025spontaneous}: This component of the LEE scales with detector volume, and has energies well below the detector threshold, forming a shot noise component in the correlated noise spectrum between phonon sensor channels. 
One possibility is that they are induced by the relaxation of lattice defects caused by the bombardment of high energy radiation.
We rescale the rate as observed on crystalline detectors.

\item LEE events in superconducting (Al) films \cite{chang2025spontaneous}: In addition to the volume-scaling LEE, there is another distinct population which do not introduce correlated signals or correlated shot noise. They have faster pulses and significantly localize around single sensor channels, suggesting that they originates from the superconducting films or the interface between the film and the bulk of the detector. 
Since the amorphous target will have much smaller volume than the crystal targets used in existing detectors, the film LEE background likely dominates the sensor resolution. 

The film LEE will also contribute background events above the detector threshold. 
This background can be suppressed by reading out one target strip with two sensors on the two opposite ends (see \cref{fig:conceptualDesign}), as the energy partition in the two channels will differ between events that originate from the Al films and the bulk of the detector.
\end{enumerate}

The above considerations motivate a total volume of 0.5 mm $\times$ 2 mm $\times$ 2 $\mu$m with the geometry shown in \cref{fig:conceptualDesign}, which corresponds to a target mass of roughly 4 $\mu$g. 
Amorphous silicon nitride and silicon-dioxide are chosen as example target materials. A $2\,\mu\text{m}$ thin amorphous membrane is patterned into isolated target strips of $500\, \mu$m by $100\, \mu$m on a Si substrate. 

To prevent the bulk background events originating from the Si substrate, it is etched away from under the amorphous membrane with the deep reactive ion etching technique.
On each target strip, there is a TES on each end. To increase the phonon collection efficiency, the TESs are coupled to 600 nm thick Al phonon absorbers with area $2500\,\mu \text{m}^2$.
Such quasiparticle trapping design has been realized in CDMS and TESSERACT detectors \cite{saab2000design,TESSERACT:2025tfw}.
Suspended membrane designs like the one proposed have already been reliably fabricated as large scale far-infrared/millimeter wave detector arrays \cite{irwin2005transition}.

We consider a TES in our example because of its relative simplicity and the world-leading resolution. Other possibilities include kinetic inductance devices (KIDs) and Josephson junction (JJ) based detectors such as charge parity boxes. The JJ-based detector SQUAT design proposed energy resolutions similar to the requirement here.

\begin{figure}
    \centering
    \includegraphics[width=0.95\linewidth]{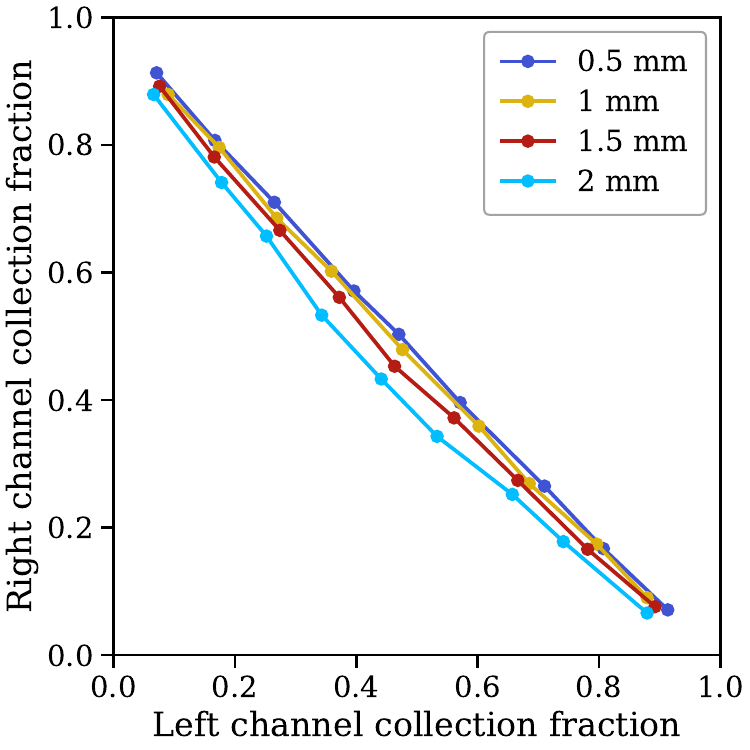}
    \caption{Phonon collection efficiency as a function of target strip length. The Al phonon collector is $25\,\mu\text{m}\times 100\,\mu\text{m}$ on each end of the strip. The high aspect ratio of the $25\,\mu\text{m}$ collection width to the $2\,\mu\text{m}$ target film thickness ensure that no phonons escape to the outside of the strip. The phonon velocity is $3.7\,\text{mm}/\mu\text{s}$, mean free path is $10\,\mu m$, and the phonon lifetime is $0.66\,\text{ms}$, assuming all phonons are at energy $2\Delta_\mathrm{Al}$. }
    \label{fig:bkg_striplength}
\end{figure}

The strip design maximizes the position dependence while maintaining sufficient phonon collection efficiency. 
Using the phonon diffusion and decay rates discussed in \cref{app:diffusion}, we carry out a simplified simulation of the collection efficiency. 
We assume that all phonons with energy less that $2\Delta_\mathrm{Al}$ are invisible to the detector and therefore lost. 
The resulting phonon collection efficiency is shown in \cref{fig:bkg_striplength} for various choices of the target strip length. At a strip length $>$ 0.5 mm, the total collection efficiency starts to degrade. 
The dots in \cref{fig:bkg_striplength} correspond to interaction locations at $10\%$ to $90\%$ of the length from right to left of the strip, in $10\%$ steps. 
This shows that it is possible to only select events from the center of the strip and reject film LEE and events from what remains of the Si substrate.
We define the fiducial volume of the detector as the middle 50\% of the strips, which results in a fiducial mass of roughly 2 $\mu$g.

\subsection{Assumed background rates}

\begin{figure}
    \centering
    \includegraphics[width=0.95\linewidth]{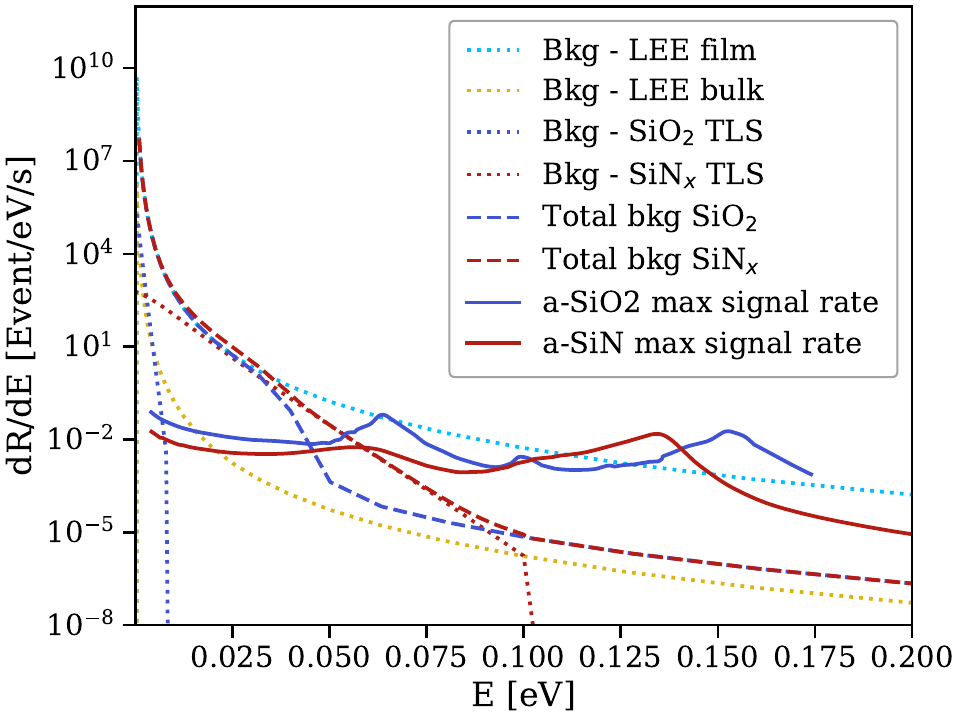}
    \caption{Estimated background and signal rates  in a 1 mm $\times$ 1 mm $\times$ 2 $\mu$m detector, for SiO$_2$ and SiN$_x$.}
    \label{fig:bkg_bkg}
\end{figure}

The rates for aforementioned background processes are shown in \cref{fig:bkg_bkg} for a 0.5 mm $\times$ 2 mm $\times$ 2 $\mu$m detector, for SiO$_2$ and SiN$_x$. 
Differences in the cut-off energy for the TLS background can be understood in terms of the different material properties of SiO$_2$ and 
SiN$_x$, as explained in \cref{app:background}.
 While the latter presents a lower total TLS relaxation rate, the high cut-off almost covers the signal region entirely. On the other hand, although SiO$_2$ has a high TLS relaxation rate below 10 meV, it contributes negligibly to the signal region. 

The bulk LEE background (yellow dotted line in \cref{fig:bkg_bkg}) is down-scaled from the measured value from data collected with 10 mm $\times$ 10 mm $\times$ 1 mm and a 10 mm $\times$ 10 mm $\times$ 4 mm crystal silicon detectors \cite{chang2025spontaneous}. 
The background power was measured to be $\sim 5$ fW/g, with an average event energy on the scale of 0.6 meV. The spectrum shape is undetermined and both exponential ($R \sim rV \exp(-E/E_0)$) and power laws ($R \sim rV(E/E_0)^{-\eta}$) are consistent with the data. 
Here we chose a power law with $n=5$ and $E_0=0.6$ meV as a pessimistic scenario, as it predicts a higher background rate in the signal region. 
The total background power above aluminum superconducting gap is normalized to $5$ fW/g.

The film LEE background (cyan dotted line in \cref{fig:bkg_bkg}) is also down-scaled from the same data set. The device in \cite{chang2025spontaneous} had a $1.6\, \text{mm}^2$ Al phonon collector area and measured roughly $1$ fW noise power, with the same characteristic energy $E_0=0.6$ meV. For simplicity, we assume the same spectral shape as the bulk LEE in our conceptual design, and scale the Al phonon collector area to 100 $\mu$m $\times$ 25 $\mu$m, as shown in Fig.~\ref{fig:conceptualDesign}, to reduce film LEE contributions. 
The film LEE is expected to inject background events above the detector threshold.
We assume that measuring the energy partition between the two channels is sufficient to reduce this rate by factor of $10^3$. 
This can be achieved by selecting events along the center diagonal region in Fig.~\ref{fig:bkg_striplength}, similar to the event classification demonstrated in \cite{pyle2006quasiparticle}. 
Under this assumption, we find that the TLS relaxation is the dominant background.

\subsection{Detector resolution}
To detect $>50$ meV dark photon signals, the detector's energy resolution should be as low as 10 meV. Assuming a 50\% phonon collection and quasiparticle trapping efficiency, the total sensor (TES) resolution should achieve $\sim$ 5 meV.
In order to measure the event location, two sensors are required, and their noise will be added quadratically, implying that each of them should have roughly $\sim 5\,\text{meV}/\sqrt2\approx 3\,\text{meV}$ resolution.

The quasiparticle Fano noise is considerable at this energy level. It is due to the number fluctuations when converting the total energy to the energy quanta, broken Cooper-pairs in this case. 
A $50$ meV energy deposition on average breaks $140$ Cooper-pairs in Al, assuming a pessimistic scenario of a Fano factor of one and the 50\% phonon collection efficiency, the total sensor resolution from Fano noise is $\sqrt{140\times50\%}\times2\Delta_{Al}=3$ meV.

The TES resolution is driven by two sources: first is the intrinsic resolution associated to the operation principle, second is the external phonon shot noise from the subthreshold high-rate backgrounds. The intrinsic resolution is proportional to the square root of its volume. The dominant noise in a TES is the thermal fluctuation noise. A current world-leading TES has $0.1$ eV intrinsic resolution with $10^{6}\,\mu \mathrm{m}^3$ active volume \cite{chang2025spontaneous}. 
A resolution around 5 meV can in principle be achieved by reducing the sensor active volume by a factor of 400. 
Such a small TES has $O(10)$ aW level saturation power, running it requires careful noise reduction along the readout wires and in the sample box.

\begin{figure}
    \centering
    \includegraphics[width=0.95\linewidth]{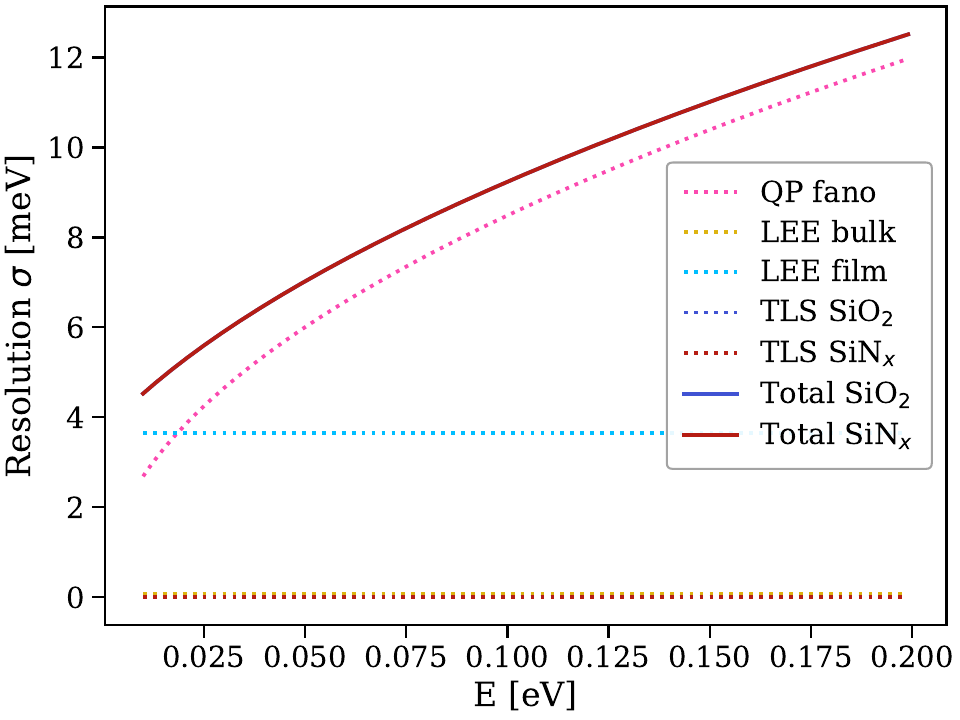}
    \caption{Expected energy resolution of the conceptual detector. The two total resolution curves are visually identical, both driven by the quasiparticle fano noise and the sensor film LEE shot noise.
    }
    \label{fig:bkg_resolution}
\end{figure}

On the other hand, the 0.1 eV resolution in current TESSERACT sensor is known to be limited by phonon shot noise \cite{chang2025spontaneous}, which can be estimated by calculating the variance of background energy in the time scale of a signal pulse $\tau_s$
\begin{equation}\label{eq:shotnoise}
    \sigma_\mathrm{ph}^2 = \tau_s\int_{2\Delta} E^2 \frac{dR_\mathrm{bkg}}{dE}dE \sim \tau_s E_0 P_\mathrm{bkg}
\end{equation}
Taking $\tau_s\approx$ 1 ms, the resulting resolution from each of the prior mentioned backgrounds is shown in \cref{fig:bkg_resolution}.
The TLS relaxation background and bulk LEE contribution to \cref{eq:shotnoise} are negligible in a 4 $\mu$g detector. The film LEE does a priori contribute significantly, which limits the area of the Al film if one seeks to maintain the desired $\sim $ 5 meV resolution. 
This why the area of the Al film was limited to $2500\,\mu\text{m}^2$, which corresponds to a $\sim 3.8$ meV resolution contribution from film LEE through \eqref{eq:shotnoise}.

\subsection{Sensitivity projection}
The dark photon creates a mono-energetic signal, and as such the detector's sensitivity directly depends on the energy resolution and the background rate.
As shown in \cref{fig:bkg_bkg}, the signal curves represent the maximum differential signal rate across all DM masses, assuming the resolution shown in \cref{fig:bkg_resolution}. (Note that the expected signal follows a Gaussian distribution of unknown mean value, not the curve in \cref{fig:bkg_bkg}.) 
With the optimum interval statistical method \cite{Yellin:2008da}, an exclusion sensitivity can be calculated for the conceptual detector, as shown in \cref{fig:projected_sensitivity}.

\section{Background from TLS decays \label{app:background}}
In this appendix we estimate the background from decay of excited TLSs in the target. 
We first review the relevant features of the TLS, before calculating their contribution to the background rate.

\subsection{Two-level systems}

The aforementioned measurements by Zeller and Pohl revealed a curious linear temperature dependence of the thermal conductivity at low temperatures~\cite{PhysRevB.4.2029}. Anderson, Halperin, and Varma~\cite{Anderson01011972} and Phillips~\cite{Phillips:1972sbs} independently explained this behavior by postulating an ensemble of two-level systems (TLS), which arise from the disorder within the amorphous material. Resonant scattering of phonons from TLS limits the phonon mean free path at low energies, in contrast to crystals where long-wavelength modes are insensitive to structure on the scale of the lattice spacing, thus modifying the thermal transport properties. For a more recent review of TLS, see \cite{yu2021twolevelsystemstunnelingmodel}.

\begin{figure}
    \centering
    \includegraphics[width=0.85\linewidth]{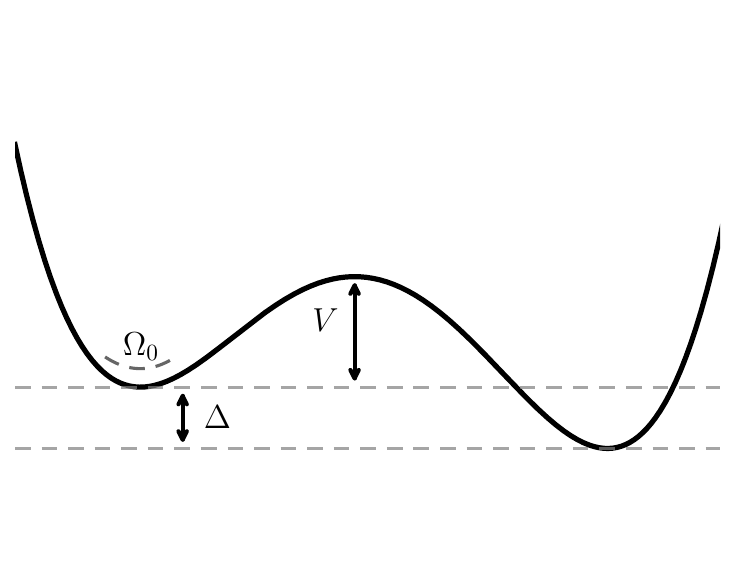}
    \caption{Schematic representation of a two-level system. $\Omega_0$ is the characteristic frequency of the minimum.}
    \label{fig:doublewell}
\end{figure}

Physically, the TLS are thought to correspond to atoms or small groups of atoms that occupy two nearly degenerate configurational states separated by an energy barrier. Each system is modeled as a one-dimensional asymmetric double-well potential (see \cref{fig:doublewell}), characterized by a barrier height $V$ and an asymmetry energy $\Delta$.
These parameters follow a broad and roughly uniform distribution \cite{PhysRevB.84.174109}. Transitions between the two configurational states can be induced by phonon emission or absorption, with the relevant phonon energy satisfying $\omega \approx \Delta$.

At sufficiently low temperatures, the TLS must relax via quantum tunneling, and the corresponding rate is therefore exponentially sensitive to the barrier height $V$. The wide distribution of barrier heights implies that some TLS decay immediately, while others have lifetimes long enough to inject energy into the experiment days, weeks, or even years after the experiment was cooled down to cryogenic temperatures, thus producing an irreducible background. We attempt to estimate this background in the remainder of this appendix.

\subsubsection{TLS Hamiltonian and relaxation rate \label{sec:TLSrelax}}

Following reference \cite{PhysRevB.84.174109}, the TLS Hamiltonian and the phonon induced perturbations are 
\begin{align}
    H_0 & =\frac{1}{2}\left(\Delta \sigma_z - \Delta_0 \sigma_x\right)\\
    H_1 &= \gamma \epsilon \sigma_z
\end{align}
where $\Delta_0$ is the tunneling matrix element, $\gamma$ the phonon-TLS coupling, $\epsilon$ the strain field and $\sigma_{x,z}$ are Pauli matrices. The energy eigenstates of $H_0$ have splitting 
\begin{equation}
    E\equiv \sqrt{\Delta^2+\Delta_0^2}\approx \Delta. 
\end{equation}

Quantizing the strain field, the matrix element for phonon emission (or absorption) is
\begin{equation}
    \mathcal{M} = \langle \bfq,\alpha, 1| H_1 | 0 \rangle = i q \sqrt{\frac{\hbar}{2\rho \mathcal{V} \omega }}\gamma \frac{\Delta_0}{E}
\label{eq:phonon_matrix_element}
\end{equation}
with $q, \omega$ the phonon momentum and energy, $\rho$ the mass density and $\mathcal{V}$ the volume of the sample. This matrix element is the same for absorption, stimulated emission and decay. 

The tunneling matrix element $\Delta_0$ is given in the WKB approximation by:
\begin{equation}\label{eq:Delta0}
    \Delta_0 = \frac{\hbar\Omega_0}{\pi}\left(\sqrt{\Lambda + 1} + \sqrt{\Lambda}\right)\exp\left(-\sqrt{\Lambda^2 + \Lambda}\right),
\end{equation}
where $\Lambda = 2V/(\hbar\Omega_0)$, and $\Omega_0$ is the zero-point frequency of the potential well (see \cref{fig:doublewell}).

Applying Fermi's golden rule with $E = \omega$ yields the decay rate
\begin{equation}\label{eq:TLSlifetime}
\tau_d^{-1} = \frac{\Gamma_{dec}}{\hbar} = \sum_{a=l,t}\left(\frac{\gamma_a^2}{v_a^5}\right)\frac{\Delta_0^2 \omega}{2\pi \hbar^4 \rho}.
\end{equation}
where the sum runs over longitudinal ($\ell$) and both transverse ($t$) modes with sound velocities $v_a$. 
In a cryogenic environment, the 10 mK ambient temperature ensures that the energy of the thermal phonons is much less than the $~$meV-scale phonons. The lifetime due to direct phonon relaxation is then dominated by the decay rate and we can neglect stimulated emission and absorption.

\begin{figure}[tbp]
 \centering
    \includegraphics[width=\linewidth]{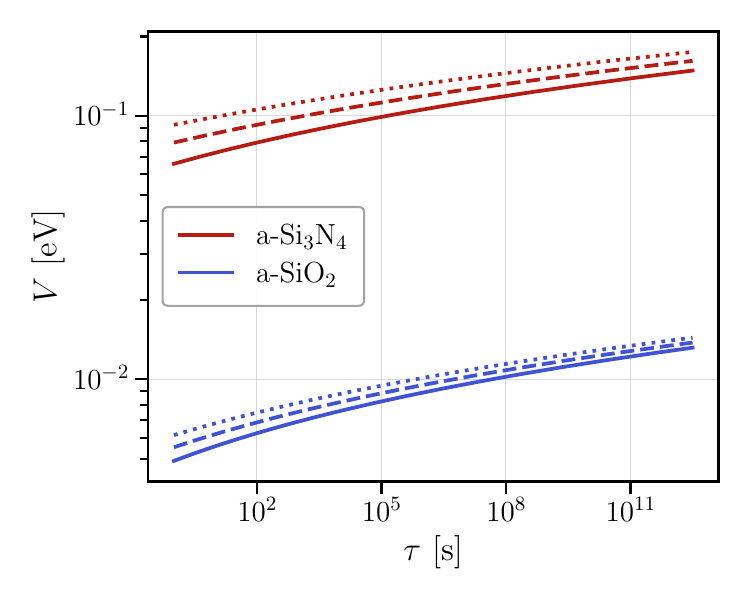}

\caption{Contours of potential barrier height $V$ and decay lifetime $\tau$ for fixed phonon energy $\omega$. The red (blue) lines are for Si$_3$N$_4$ (SiO$_2$). Dotted, dashed and solid lines correspond to $\omega=10^{-1}\mathrm{eV}, 10^{-3}\mathrm{eV}$ and $10^{-5}\mathrm{eV}$ respectively. The lifetime is calculated assuming $T = 0\,\mathrm{K}$ and is determined solely by quantum tunneling processes using the WKB approximation. Due to the exponential dependence on barrier height, a wide range of lifetimes (spanning many orders of magnitude) corresponds to small variations in $V$.}
\label{fig:TLS_V_combined}
\end{figure}

This lifetime is shown in \cref{fig:TLS_V_combined} for amorphous $\text{SiO}_2$ and $\text{Si}_3\text{N}_4$; the different lifetimes are due to the different $\Omega_0$ in both materials (see \cref{tab:TLSparameters}).  
As evident from \eqref{eq:TLSlifetime}, the lifetime is inversely proportional to $\omega$ and depends extremely strongly on the barrier height $V$.
In practice, this means that only a very small range in $V$ contribute to the background rate:
For the smaller $V$ one can simply wait a short period for the TLS to decay away exponentially before taking data for a dark matter search.
Once the TLS lifetime exceeds the duration of the experiment, for example $\sim 1$ year, the TLS contributes to the background rate proportional to $\sim 1/\tau_{d}$. The contributions from the TLS with higher $V$ are therefore also exponentially suppressed.
The background rate from TLS relaxation will thus be determined only by the number of occupied TLS with $V\approx 0.01$ eV ($V\approx 0.1$ eV) for $\text{SiO}_2$ ($\text{Si}_3\text{N}_4$).
Next, we need to determine what fraction of the TLS have barriers of this order, and what their occupation number is at the start of the experiment.

Before calculating this, we mention a few caveats to our estimate of the lifetime. Firstly, the matrix element in \cref{eq:phonon_matrix_element} is strictly speaking only valid below the Ioffe-Regel transition transition, while a priori we are also interested in higher frequencies.
Moreover, in more sophisticated TLS models it has been argued that the tunneling matrix element is renormalized downwards by a polaron effect \cite{carruzzo2020phonon}.
As we will see below, the strong dependence on the barrier height $V$ will render even very large multiplicative uncertainties in \cref{eq:phonon_matrix_element} quantitatively irrelevant for the purpose of dark matter detection.

\subsubsection{TLS probability distribution}

To compute the total background from TLS relaxation, we must specify the number density and distribution of TLS parameters. We assume that the asymmetry $\Delta$ is uniformly distributed and that the barrier heights $V$ follow a Gaussian distribution:
\begin{equation} \label{eq:TLSdistribution}
    P(\Delta, V) = \frac{2\bar P}{\hbar \Omega_0}\exp\left[-\frac{(V-V_0)^2}{\sigma_0^2}\right].
\end{equation}
This model is characterized by four parameters: the potential well energy spacing $\Omega_0$, the TLS number density normalization $\bar{P}$, and the parameters $V_0$ and $\sigma_0$ describing the Gaussian distribution of barrier heights. 
These parameters can be determined experimentally by fitting to measurements of mechanical dissipation, specific heat, and thermal conductivity as functions of temperature \cite{PhysRevB.84.174109}. The cited work also includes the contribution of localised vibrational modes (``Einstein oscillators") to the dissipation in the material above the Ioffe-Regel limit.
These additional excitations only affect the high frequency spectrum and are too short-lived to constitute a meaningful background in a dark matter experiment.

 While reference \cite{PhysRevB.84.174109} integrated the TLS probability distribution up to $V_{\mathrm{max}}=V_0+6\sigma_0$ and $E_{\mathrm{max}}=2V$, we find our background estimates are insensitive to the tail end of the TLS distribution.

\begin{table}[h!]
\centering
\begin{tabular}{lcc}
\hline\hline
\textbf{Quantities} & \textbf{SiO$_2$} & \textbf{Si$_3$N$_4$} \\
\hline
$\rho$ ($10^3$ kg/m$^3$) & 2.2 & 3.18 \\
$v_L$ ($10^3$ m/s) & 5.8(L) 3.75(T) & 11.17 \\
$\bar{P}$ ($10^{45}$/Jm$^3$) & 0.16 & $\sim$0.39 \\
$\gamma$ (eV) & 2.24(L) 1.73(T) & 5.6 \\
$\hbar \Omega_0$ (K) & 12 & 130 \\
$V_0$ ($\times10^4$ K) & 0 & 3.05 \\
$\sigma_0$ ($\times10^3$ K) & 0.445 & 7.5 \\
\hline\hline
\end{tabular}
\caption{TLS model parameters for SiO$_2$ and Si$_3$N$_4$, obtained from \cite{PhysRevB.84.174109}}
\label{tab:TLSparameters}
\end{table}

The TLS model parameters used in our background estimates are summarized in Table~\ref{tab:TLSparameters}. For Si$_3$N$_4$, we adopt values corresponding to a high-stress film. The analysis of Ref.~\cite{PhysRevB.84.174109} also provides parameters for stress-relieved Si$_3$N$_4$ and for SiN${_{1.15}}$. Relative to the high-stress sample in the table, the stress-relieved Si$_3$N$_4$ and SiN$_{1.15}$ samples exhibit an approximately $15\%$ larger value of $\Omega_0$, while the normalization $\bar{P}$ differs by approximately an order of magnitude between the Si$_3$N$_4$ samples and SiN${_{1.15}}$.

\subsection{Background estimates}

With the TLS model established, we now estimate the background rate from TLS relaxation. 
This requires estimating the number of excited TLS at the start of the experiment.

\subsubsection{Freeze-out during cooling}

The TLS occupation fraction is given by its equilibrium distribution at the temperature at which the TLS lost contact with the thermal phonon bath. 
This in turn depends on the potential barrier $V$ and the rate at which the sample is cooled.
To quantify this, we set up the kinetic equations and carry out the TLS freeze-out calculation.

We define $n_0$ and $n_1$ as the ground and excited state occupancy fractions, with the constraint that $n_1=1-n_0$. The kinetic equations are
\begin{align}
\frac{d n_0}{dt} &= \Gamma_\downarrow n_1(t) - \Gamma_\uparrow n_0(t)\\
\frac{d n_1}{dt} &= \Gamma_\uparrow n_0(t) - \Gamma_\downarrow n_1(t)
\end{align}
with $\Gamma_{\uparrow,\downarrow}$ the total excitation and de-excitation rates. 
This can be written more compactly by defining \mbox{$\Delta n \equiv n_1 -n_0$} which obeys
\begin{equation}\label{eq:kinDeltan}
\frac{d \Delta n}{dt} = -\overline\Gamma  (\Delta n(t)- \Delta n_{eq})
\end{equation}
with $\bar\Gamma\equiv \Gamma_{\downarrow}+\Gamma_{\uparrow}$ and $\Delta n_{eq} \equiv\frac{\Gamma_{\uparrow}-\Gamma_{\downarrow}}{\Gamma_{\uparrow}+\Gamma_{\downarrow}}$. $\Delta n$ relates to the $n_{0,1}$ through
\begin{align}
n_0&=\frac{1-\Delta n}{2}\\
n_1&=\frac{1+\Delta n}{2}.
\end{align}

Assuming that the rates are thermal, they obey
\begin{equation}
    \frac{\Gamma_{\uparrow}}{\Gamma_{\downarrow}}= e^{-\beta \omega}
\end{equation}
with $\omega$ the energy splitting of the TLS and $\beta=1/k_bT$ the inverse temperature. Then the equilibrium solution is
\begin{equation}
    \Delta n^{eq} = -\tanh\left(\frac{\beta \omega}{2}\right)
\end{equation}
which translates to 
\begin{align}
n_0^{eq} &=\frac{1}{1+e^{-\beta\omega}}\\
n_1^{eq} &=\frac{1}{1+e^{\beta\omega}}
\end{align}
The latter corresponds to the Fermi-Dirac distributions with zero chemical potential. 

During the cooling down of the experiment towards cryogenic temperatures, the most efficient way to excite or relax the TLS is through a local thermal fluctuation.\footnote{For $\beta V\ll 1$ phonon assisted tunneling becomes more important, though we verified that this process is not efficient enough to maintain thermal equilibrium.}  
This is called the Arrhenius process, with rates
\begin{align}
\Gamma_{\uparrow} &= \frac{\Omega_0}{2} e^{-\beta V} \label{eq:arrup}\\
\Gamma_{\downarrow}\label{eq:arrdown} &= \frac{\Omega_0}{2} e^{-\beta (V-\omega)}. 
\end{align}
with $\Omega_0$ the typical zero-point frequency of the double well potential.
Once the temperature drops below the barrier height, $\Gamma_{\uparrow} \ll \Gamma_{\downarrow}$ and the Arrhenius process continues to deplete the TLS as long as $\Gamma_{\downarrow}$ is still sufficiently efficient.

\subsubsection{Results}

\begin{figure}[tbp]
    \centering
    \includegraphics[width=\linewidth]{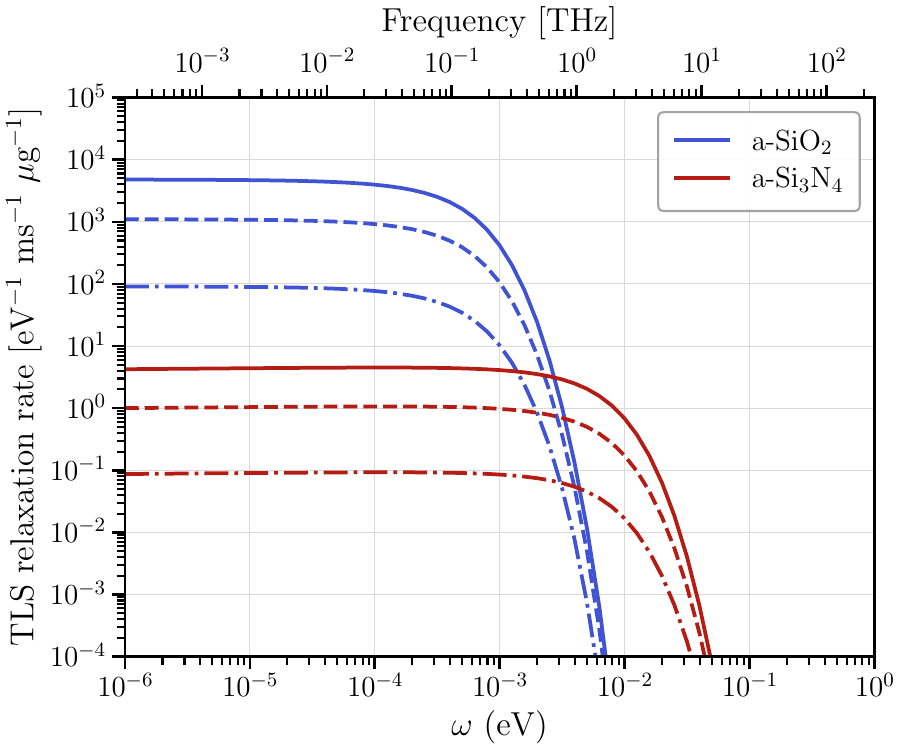}
    \caption{The TLS relaxation rate from quantum tunneling into phonons of energy $\omega$. The solid, dashed, and dot-dashed lines correspond to 1 week, 1 month, and 1 year after the system is initially cooled, respectively.  The material is assumed to be cooled from an initial temperature of $\sim 300\,\mathrm{K}$ with a cooling rate given by eq.~\ref{eq:coolrate}, such that Arrhenius relaxation processes (eqs.~\ref{eq:arrup}, \ref{eq:arrdown}) lead to a high-energy cutoff of the excited TLS distribution. 
    }
    \label{fig:relaxation_rates}
\end{figure}

Starting at room temperature $(T=300\,\text{K})$, we assume a cooling time of 1 day, defined by 
\begin{equation}\label{eq:coolrate}
T(t) = \text{Max}\left[300 \,\mathrm{K} \times  \exp\left(\frac{-t}{\, \mathrm{hour}}\right),\,10\,\text{mK}\right]
\end{equation}
effectively rendering the rates time-dependent. 
We can then numerically solve \eqref{eq:kinDeltan} and fold in \eqref{eq:TLSdistribution} to obtain the number of TLS with a given $V$ and $\omega$ that are excited at the start of the experiment. While the relic TLS density is further lowered by cooling more slowly, improvements are minor.  
Concretely, we have tested different cooling profiles, for example a faster or slower exponential cooling profile, linear cooling, and also starting from a higher temperature $\sim 1000\,\mathrm{K}$ which is approximately the formation temperature of the material. We find small differences in the relic population of TLS, since the dependence on the cooling profile is only logarithmic (as motivated in the next sub-section).

After cooling, those excited TLS states can still decay via quantum tunneling processes described by \eqref{eq:TLSlifetime} and \cref{fig:TLS_V_combined}. 
The resulting background event rate is shown in \cref{fig:relaxation_rates}. 
At low frequencies, the TLS background is large and approximately flat in frequency. 
The amplitude is several orders of magnitude smaller for $\mathrm{Si_3N_4}$ compared to $\mathrm{SiO_2}$, consistent with the fact that $\mathrm{Si_3N_4}$ is known to exhibit lower (but nevertheless significant) TLS noise in GHz devices such as superconducting qubit circuits and microwave resonators~\cite{Martinis_2005,PhysRevB.84.174109}. 

The high-energy cutoff in \cref{fig:relaxation_rates} is critical for the potential viability of an amorphous sensor as a dark matter detector, and effectively sets a fundamental lower threhold. 
It arises from the TLS freeze-out and is determined primarily by the relevant range of TLS barrier heights (V), which in turn depends most sensitively on the potential well frequency $\Omega_0$ of the material. 
Incidentally, a measurement of this cutoff could precisely determine the value of $\Omega_0$, which may be relevant for better understanding TLS dynamics across a broad range of devices and frequencies. We adopt $\text{Si}_3\text{N}_4$ as a fiducial reference; however, Ref.~\cite{PhysRevB.84.174109} reports comparable $\Omega_0$ values for both $\text{SiN}_{1.15}$ and $\text{Si}_3\text{N}_4$. We therefore expect similar behavior across $\text{SiN}_x$ stoichiometries.

As explained above, the matrix element \eqref{eq:phonon_matrix_element} was obtained in the WKB approximation and is strictly speaking not valid for energies above the Ioffe-Regel limit, where the mean free path is smaller than the wavelength of the mode \cite{Allen01011999}. 
This transition occurs for $\omega$ around the exponential cut-off in \cref{fig:relaxation_rates}, as explained in the analytical treatment below.
We however expect that exponential dependence of the tunneling amplitude on the barrier height $V$ should hold regardless, and thus not qualitatively alter the conclusions. 

\subsubsection{Analytical treatment}
To conclude, we provide an approximate analytic argument for the exponential drop-off in \cref{fig:relaxation_rates}.
As the sample is cooled, various TLS drop out of thermal equilibrium with the phonons. Those with the highest $V$ drop out first, as is evident from \eqref{eq:arrup} and \eqref{eq:arrdown}. This happens once the TLS-phonon interaction rates become slower than the change in temperature:
\begin{equation}
    \bar \Gamma \lesssim \frac{|dT/dt|}{T}
\end{equation}
For the cooling profile in \eqref{eq:coolrate} this inequality is saturated for
\begin{equation}
\frac{\Omega_0}{2}e^{-\beta (V-\omega)} \approx \frac{1}{\text{hour}}
\end{equation}
where we assumed the temperature is sufficiently low the $\Gamma_\uparrow\ll \Gamma_{\downarrow}$.
From this we derive
\begin{equation}
    \beta V \approx \log\left(\frac{1}{2}\Omega_0 \times \text{hour}\right) \approx 34
\end{equation}
where we further took $V-\omega \approx V$.
Next, we can solve for $\beta$ in the expression in the equilibrium concentration for $n_1$
\begin{equation}
    n_1^{eq} =\frac{1}{1+e^{34\frac{\omega}{V}}}\approx e^{-34\frac{\omega}{V}}.\label{eq:exponentialcondition}
\end{equation}
We see that the left over number density of the excited TLS is exponentially suppressed for $\omega/V\gtrsim 34$.
In \cref{fig:TLS_V_combined} we saw that the lifetime grows exponentially and the TLS are effectively stable for $V\gtrsim 0.01$ eV ($V\gtrsim 0.1$ eV) in $\text{SiO}_2$ ($\text{Si}_3\text{N}_4$).
Together with \eqref{eq:exponentialcondition} this explains why the background rate starts to drop exponentially around $\omega \gtrsim 1$ meV ($\omega \gtrsim 10$ meV) in $\text{SiO}_2$ ($\text{Si}_3\text{N}_4$).

We can further understand this in terms of $\Omega_0$: The background rate is dominated by TLS with lifetime that is of the order of the time we let TLS decay away before taking data. 
Taking this to be $\sim$ day, from \eqref{eq:TLSlifetime} we extract
\begin{equation}
    \Lambda = V/\hbar\Omega_0 \approx 20
\end{equation}
with only a logarithmic dependence on the other parameters in \eqref{eq:TLSlifetime}. 
Plugging this into \eqref{eq:exponentialcondition}, we find that the exponential drop in the background occurs for 
\begin{equation}
    \omega \gtrsim \hbar\Omega_0.
\end{equation}
This explains why this occurs for an $\omega$ that is about an order of magnitude larger in $\text{Si}_3\text{N}_4$ as compared to $\text{SiO}_2$.
To leading order, the remaining parameters in \cref{tab:TLSparameters} only impact the height of the plateau in \cref{fig:relaxation_rates}. 

The value of $\Omega_0$ is determined from a fit of the TLS model to measurements of the temperature dependence of the acoustic dissipation. It correlates most strongly with an observed rise in the acoustic dissipation, which occurs around T=10 K for $\text{SiO}_2$ \cite{PhysRevB.84.174109}.  
As shown in \cite{PhysRevB.84.174109}, an order one variation of $\Omega_0$ qualitatively fails to describe the data, suggesting that $\Omega_0$ can be extracted fairly reliably. 
That said, the TLS with low barriers dominate the acoustic dissipation, while the high barrier TLS are relevant in a dark matter experiment. 
Implicitly, we are therefore assuming that the $\Omega_0$ and $V$ distributions are uncorrelated. 
If this assumption is invalid, this could qualitatively influence the background model that let to \cref{fig:relaxation_rates}. 
As far as we are aware, constructing a prototype sensor is currently the best way to validate this assumption.

\end{document}